\documentclass[a4paper,11pt]{article}

\usepackage[margin=2.5cm]{geometry}

\usepackage{graphicx}
\usepackage{soul}
\usepackage{caption}
\usepackage{subcaption}
\usepackage{adjustbox}
\usepackage{comment}
\usepackage{tabularx}
\usepackage{booktabs}
\usepackage{newunicodechar}
\usepackage{url}
\usepackage[utf8]{inputenc}

\providecommand{\keywords}[1]
{
  \small	
  \textbf{\textit{Keywords---}} #1
}

\begin{document}

\title{Information disorders on Italian Facebook during COVID-19 infodemic}

\author{
        Alessandro Celestini$^{1}$, Marco Di Giovanni$^{2}$, Stefano Guarino$^{3}$, Francesco Pierri$^{4}$
 \\\\
        \small $^{1,3}$Institute for Applied Computing "Mauro Picone", National Research Council, Rome, Italy. \\
        \small $^{2,4}$Department of Electronics, Information and Bioengineering, Politecnico di Milano, Milano, Italy.\\\\
        Email:  $^{1}$a.celestini@iac.cnr.it, $^{2}$marco.digiovanni@polimi.it, \\$^{3}$s.guarino@iac.cnr.it, $^{4}$francesco.pierri@polimi.it
}

\date{}

\maketitle              

\begin{abstract}
In this work we carry out an exploratory analysis of online conversations on the Italian Facebook during the recent COVID-19 pandemic. 
We analyze the circulation of controversial topics associated with the  origin of  the  virus, which involve popular targets of misinformation,  such  as  migrants  and  5G  technology.
We collected over 1.5 M posts in Italian language and related to COVID-19, shared by nearly 80k public pages and groups for a period of four months since January 2020. Overall, we find that potentially harmful content shared by unreliable sources is substantially negligible compared to traditional news websites, and that discussions over controversial topics has a limited engagement w.r.t to the pandemic in general. Besides, we highlight a ``small-worldness" effect in the URL sharing diffusion network, indicating that users navigating through a limited set of pages could reach almost the entire pool of shared content related to the pandemic, thus being easily exposed to harmful propaganda as well as to verified information on the virus.
\end{abstract}


\section{Introduction and related work}
The spread of a novel coronavirus (COVID-19) in the past months has changed in an unprecedented way the everyday life of people on a global scale. According to World Health Organization (WHO), at the time of this writing the pandemic has caused over 3.5 M confirmed cases, with more than 250k fatalities globally speaking\footnote{\url{https://covid19.who.int}}. Italy, in particular, has been one of the first European countries to be severely hit by the pandemic, as the virus spread outside China borders at the end of January, and to implement national lockdown on the 9th of March \cite{dpcm9marzo,bonaccorsi2020}. Following Italy and China, national lockdowns have been adopted by most countries around the world, drastically reducing mobility flows in order to circumvent the spread \cite{galeazzi2020}.

In relation to the emergency, the term ``infodemic" has been coined to describe the risks related to the massive spread of harmful and malicious content on online social platforms \cite{zarocostas2020fight}, as misinformation could support the spread of the virus undermining medical efforts and, at the same time, drive societal mistrust producing other direct damages \cite{zarocostas2020fight}. 
In response, several contemporary works have provided different perspectives on this phenomenon. Authors of \cite{Gallotti2020} analyzed more than 100 millions Twitter messages posted worldwide in 64 languages and found correspondence between waves of unreliable and low-quality information and the epidemic ones. Authors of \cite{Yang2020} have investigated the prevalence of low-credibility content in relation to the activity of social bots, showing that the combined amount of unreliable information is comparable to the retweets of articles published on The New York Times alone. Finally, authors of \cite{Cinelli2020} have carried out a comparative analysis of information diffusion on different social platforms, from Twitter to Reddit, finding different volumes of misinformation in different environments.

As a matter of fact, ever since 2016 US presidential elections we witnessed to a growing concern of the research community over deceptive information spreading on online social networks \cite{allcott2017,Lazer18,botnature,Pierri2019}. In Italy, according to Reuters, trust in news is today particularly low \cite{reuters2019}, and previous research has highlighted the exposure to online disinformation in several political circumstances, from 2016 Constitutional Referendum to 2019 European Parliament elections \cite{caldarelli2019,guarino2019,guarino2020characterizing,quattrociocchi2017,PierriArtoni2020}. A recent questionnaire by SOMA observatory on disinformation spreading on online social media\footnote{\url{http://www.t-6.it/report-on-the-role-of-the-information-in-the-emergency-covid-19-impacts-and-consequences-on-people-behaviors-report/}} (a project funded by the European Union) showed that people relied on official channels used by authoritative institutions in order to inform about the pandemic. Interestingly, social media were not the primary source of information during the crisis. 

In this work we adopt a consolidated strategy to label news articles at the source level \cite{Grinberg,Bovet2019,botnature,PierriScirep2020,PierriWWW20} and investigate accordingly the diffusion of different kinds of information on Facebook. We thus use the term ``disinformation" as a shorthand for unreliable information in several forms, all potentially harmful, including false news, click-bait, propaganda, conspiracy theories and unverified rumours \cite{PierriScirep2020,PierriWWW20}. We use instead the term ``mainstream" to indicate traditional news websites which convey reliable and accurate information. This approach has been mainly used for Twitter, which however exhibits a declining trend as a platform to consume online news \cite{reuters2019,PierriArtoni2020}. 
Similar to \cite{giglietto2018}, we leverage Crowdtangle platform to collect posts related to COVID-19 from Facebook public pages and groups. We use a set of keywords related to the epidemic and we limit the search to posts in the Italian language. The overall dataset accounts for over 1.5 M public posts shared by almost 80k unique pages/groups. We investigate the prevalence of reliable vs non-reliable information by analysing the domain of URLs included in such posts. In particular, we are interested in understanding how specific disinformation narratives compete with official communications. To this aim, we further specify keywords related to three different controversial topics that have been trending in the past few months, all related to the origins of the novel coronavirus: (1) the alleged correlation between COVID-19 and migrants, (2) between the virus and 5G technology, and (3) rumours about the artificial origin of the virus.

This work provides the following main contributions:
\begin{itemize}
    \item We evaluate the prevalence of COVID-19 (and related controversial topics) on the Italian Facebook, identifying the key players and the most relevant pieces of content in the information ecosystem in terms of both volume of posts and generated engagement.
    \item We study how these issues shaped the debate on Facebook, quantifying the polarization of groups and pages w.r.t. topics of discussion and measuring the respective lexical/semantic divergence.   
    \item We analyze patterns in the URL sharing network of groups/pages, observing that the majority of groups and pages interact in a ``small-world'', and discarding the hypothesis that different groups draw upon fully separate pools of web resources.
    \item We focus on the connections among URLs related to controversial topics and among groups/pages where these URLs where shared, to assess the existence of clusters induced by different sources of information and to ascertain that centrality in these networks is not correlated with high engagement.  
\end{itemize}

The outline of this paper is the following: we first describe the methodology applied, including the collection of data from Facebook, the taxonomy of news sources and controversial topics, and both text and network analysis tools; then we describe our contributions, and finally we draw conclusions and future work.

\section{Methodology}
\subsection{Facebook data collection}
We used CrowdTangle's ``historical data'' interface \cite{crowd} to fetch posts (in Italian language) shared by public pages and groups since January 1st 2020 until May 12th 2020 and containing \textit{any} of the following keywords: \textit{virus}, \textit{coronavirus}, \textit{covid}, \textit{sars-cov-2}, \textit{sars cov 2}, \textit{pandemia}, \textit{epidemia}, \textit{pandemic}, \textit{epidemic}. The tool only tracks public posts made by public accounts or groups. Besides, it does not track every public account\footnote{All pages with at least 100K likes are fully retained. For details on the coverage for pages with less likes we refer the reader to \url{https://help.crowdtangle.com/en/articles/1140930-what-is-crowdtangle-tracking}.} and does not track neither private profiles nor private groups. For each post we collected the number of public interactions (likes, reactions, comments, shares, upvotes and three second views) as well as Uniform Resource Locators (URL) attached with it. Our collection contains overall 1.59 M posts shared by 87,426 unique Facebook pages/groups. In the rest of the paper, we use ``accounts" as a shorthand to indicate the entire set of pages and groups. Data is not publicly available, but it can be provided to academics and non-profit organizations upon request to the platform.

\subsection{Mainstream and disinformation news}\label{sec:mainstream_disinfo}
To understand the prevalence of reliable vs non-reliable information we refer to the lists of news outlets compiled in \cite{PierriArtoni2020,PierriWWW20}. We use the coarse ``source-based" approach adopted in the literature \cite{botnature,Bovet2019,PierriScirep2020} to label online news articles (\emph{i.e.}, links shared in Facebook posts) in two classes according to their domain: \textbf{(1)} \textit{Disinformation} sources, which notably publish a variety of harmful information, from hyper-partisan stories to false news and conspiracy theories; \textbf{(2)} \textit{Mainstream} sources, which (mostly) provide accurate and reliable news reporting. However, this classification might not always hold since unreliable websites do share also true news, and incorrect news coverage on traditional outlets is not rare \cite{Lazer18}.
For what concerns unreliable news, we further partition the class into four distinct sets according to the geographic area: European (EU), Italian (IT), Russian (RU) and US sources.
The overall list contains 25 Italian sources for the Mainstream domain whereas for the disinformation domain we count 25 EU sources, 52 Italian sources, 13 Russian sources and 22 US sources.

\subsection{Controversial topics}\label{sec:topics}
In our analysis, we focus on three specific topics which were particularly exposed to disinformation during the infodemic\footnote{\url{https://www.newsguardtech.com/covid-19-myths/}}:
\begin{itemize}
    \item{\textbf{MIGRANTS}}: conspiracy theories that attempt to correlate the spread of the virus with migration flows. These are mainly promoted by far-right communities to foster racial hate. Some of the related keywords are: \textit{migranti}, \textit{immigrati}, \textit{ong}, \textit{barconi}, \textit{extracomunitari}, \textit{africa}.
    \item{\textbf{LABS}}: rumours that have been used as political weapons to attribute the origins of the pandemic to the development of a bioweapon to be used by China and/or to undermine the forthcoming U.S. presidential elections. Some of the related keywords are: \textit{laboratorio}, \textit{ricerca}, \textit{sperimentazione}.
    \item{\textbf{5G}}: hoaxes that can be summarized in two main streams, those claiming that 5G activates COVID-19 and those that deny the existence of the novel coronavirus and attribute its symptoms to reactions to 5G waves. Both lines are obviously false and not supported by scientific evidence. Some of the related keywords are: \textit{5g}, \textit{onde}, \textit{radiazioni}, \textit{elettromagnetismo}.
\end{itemize}
A complete list of keywords for each topic is available in the Appendix.
For sake of simplicity, we will refer to an account as a ``MIGRANTS'' account (and likewise for the other topics) if the account shared at least $N=2$ posts which contain a keyword in the related list; the same holds for URLs if the associated post contained a keyword matching the related topic.
Finally, we will denote any account or URL as ``controversial'' if it is related to at least one of the three topics.
In Table~\ref{tab:stats_accounts} we show a breakdown of the dataset in terms of posts and accounts. Note that the number of accounts is lower due to a preprocessing step described in the following paragraph.

\begin{table}[!t]
\centering
\begin{adjustbox}{width=\textwidth}
\begin{tabular}{l|ccc|cc|c}
                      & \textbf{5g}    & \textbf{Labs}   & \textbf{Migrants}   & \textbf{Intersection} & \textbf{Union} & \textbf{Total}   \\ \hline
\textbf{Posts}                 & 10937 (0.7\%) & 25695 (1.6\%) & 38486 (2.4\%) & 39 (0.024\%)           & 72440 (4.6\%) & 1588536 \\
\textbf{Accounts}    & 5493 (9.7\%)  & 7076 (12.5\%)  & 11238 (19.9\%) & 1958 (3.5\%)         & 15865 (28.8\%) & 56436   \\
\hline
\textbf{Groups' Posts} & 5817 & 15278 & 21135 & 31 & 40175 & 715104 \\
\textbf{Groups} & 3194 & 4129 & 6571 & 1232 & 9007 & 28721 \\
\hline
\textbf{Pages' Posts} & 5120 & 10417 & 17351 & 8 & 873432 & 873432 \\
\textbf{Pages} & 2299 & 2947 & 4667 & 726 & 6858 & 27715\\
\hline
\end{tabular}
\end{adjustbox}
\caption{Number of posts and accounts (groups and pages) for each controversial topic, and altogether.}\label{tab:stats_accounts}
\end{table}

\subsection{Text analysis}\label{sec:text_analysis}

We cleaned and pre-processed posts' textual content as follows. Firstly, strings are lower cased and URLs, punctuation, emojis and Italian stop words (collected from \textit{spacy} Python library) are removed. We also remove words related to the COVID-19 as they act as stop word for our analysis. 
Then, we tokenize texts using \textit{nltk} Python library~\cite{nltk}, and we remove tokens shorter than 4 or longer than 20 characters. Finally we embed accounts into vectors using Tf-Idf statistic. 
We firstly group tokens by account. To reduce noise effects we remove accounts with only 1 post and accounts with less than 20 tokens in total, obtaining 56,436 accounts from an original amount of 87,426. 
We compute the Tf-Idf of the cleaned strings, neglecting tokens that appeared less than 5 times in the whole corpus. 
Finally, for each account we obtain a sparse 137,901-dimensional embedding vector. 

\subsection{Network analysis}\label{sec:network_methodology}
We leverage tools from network science \cite{barabasi,newman} in order to investigate the diffusion network of content shared on Facebook.
We use a bipartite graph formulation to link together accounts and URLs.
Precisely, we draw an un-directed weighted edge between an account $a$ and an URL $u$ if and only if $u$ was shared at least once on/by $a$; the weight $w_{(a,u)}$ of the edge $(a,u)$ counts how many times $u$ was shared on/by $a$.
This graph has 983,582 vertices (78,760 accounts and 904,822 URLs) and 1,374,921 edges.
We then focus on the controversial bipartite graph defined as the subgraph of the accounts-URLs graph induced by controversial URLs.
This subgraph has 55,411 vertices (18,681 accounts and 36,730 URLs) and 81,707 edges.
Finally, we consider the two graphs of controversial URLs and controversial accounts obtained by projecting the giant component of the aforementioned controversial bipartite graph upon the two layers of URLs and accounts, respectively.
In other words, URLs are connected if and only if there is at least one account that shared both, and the edge is weighted by the number of such accounts.
Analogously, accounts are connected if and only if there is at least one URL that they both shared, and the edge is weighted by the number of such URLs.
In all cases, we focus on the giant connected component of such graphs, as described in Section~\ref{sec:network_analysis}.

\section{Descriptive statistics}
\subsection{Posts, reactions and news articles}

\begin{figure}[!t]
    \centering
    \includegraphics[width=\linewidth]{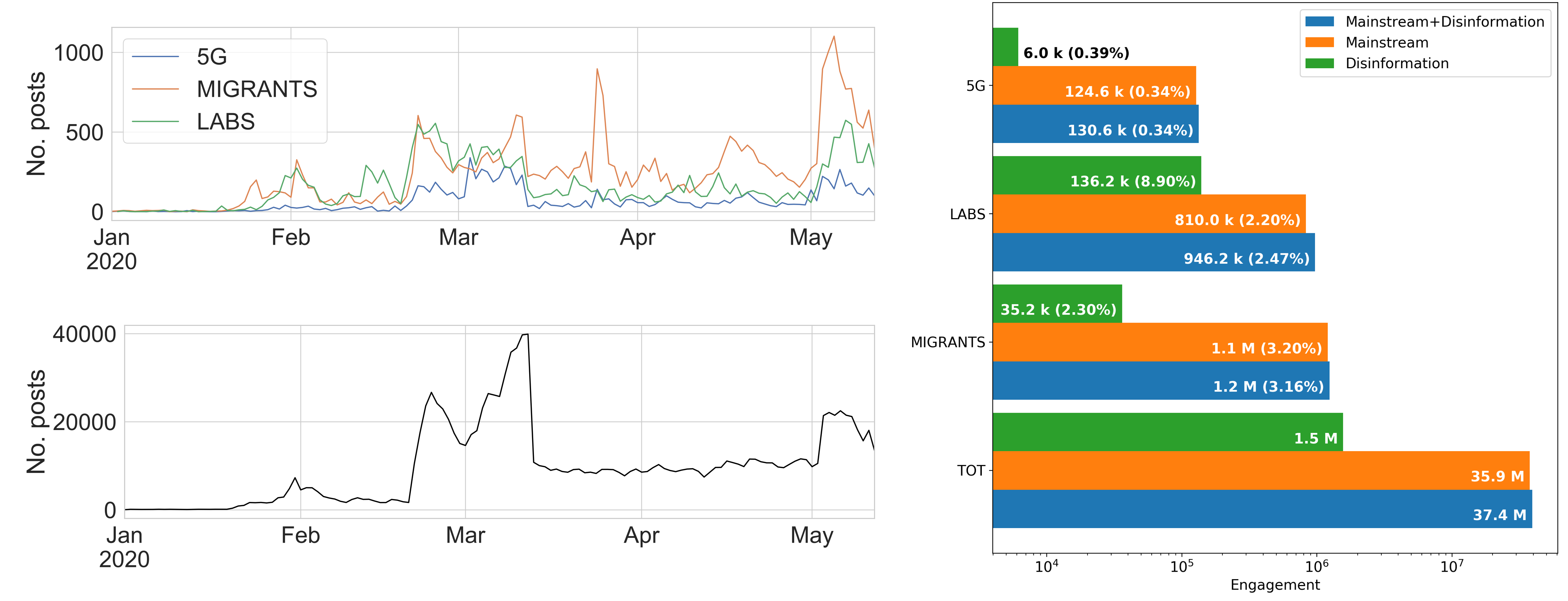}
    \caption{\textit{(left)} Time series of the daily number of posts, total and per topic. \textit{(right)} Total engagement generated by news articles for Mainstream (orange), Disinformation domains (green) and both (blue), for each topic and altogether.}
    \label{fig:prevalence_ts}
\end{figure}

We first inspect the prevalence of COVID-19 in online conversations by showing the time series of daily posts on Facebook in Figure~\ref{fig:prevalence_ts} (left). We observe a general increase in the overall volume (top), with a few spikes at the end of January (when China imposed lockdown), at the end of February (when the virus was first diagnosed in Italy), at mid March (when lockdown was applied in Italy) and at the beginning of May (when restrictions have been lifted). We provide in the Appendix the time series of daily engagement (Fig. A1), which has similar characteristics although with a different order of magnitude (up to $10^{19}$). For what concerns controversial topics (bottom), we immediately see that volumes are negligible w.r.t general conversations (the same holds for daily reactions, which are 2 order of magnitude smaller); also, they are quite aligned in time but they do exhibit spikes of their own, which are most likely related to real world events (for instance sabotages of 5G antennas in several countries).

For what concerns the diffusion of URLs, we inspect most popular domains by focusing on their total engagement. In particular, in the Top-10 ranking of domains we encounter websites which are all related to Italian Mainstream newspapers, with the exception of Facebook and YouTube which are the 2 most shared domains. When we focus on the Top-10 ranking of news websites we observe what follows (see also figures in Appendix):
\begin{itemize}
    \item Italian Mainstream newspapers generated from 2 to 6 M reactions;
    \item IT disinformation outlets generated no more than 500k reactions each; The top-3 are a generic untrustworthy website (``silenziefalsita.it"), the far-right website ``ilprimatonazionale.it"  and a law enforcement fan club (``sostenitori.it")
    \item only one RU website ``it.sputniknews.com" (which is technically in Italian language) generated more than 100k reactions, whereas the others had a negligible engagement;
    \item EU and US sources did not receive much attention (most engaged sources did not exceed 3k reactions).
\end{itemize}
Similar considerations hold also when analyzing the ranking by number of posts.
Overall, as shown in Figure~\ref{fig:prevalence_ts} (right), unreliable sources had a limited yet not negligible amount of engagement (1.5 M) compared to news websites which convey reliable information (35.9 M).
Finally it is interesting to notice that Disinformation sources generate relatively more engagement for the LABS topic, while the MIGRANTS topic was the most discussed among Mainstream news websites. 

\subsection{Accounts' characteristics}\label{sec:accounts}
\begin{figure}[!t]
    \centering
    \includegraphics[width=\linewidth]{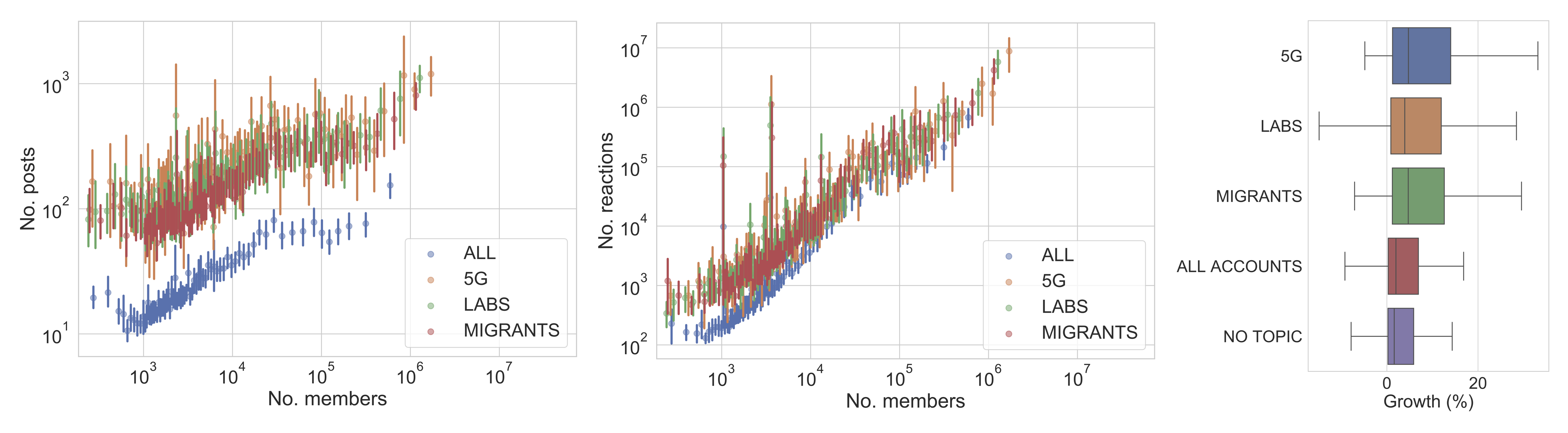}
    \caption{Scatter plot of the number of members \emph{vs.} number of reactions \textit{(left)} and number of posts \textit{(center)} for each account. Points are grouped in 100 bins to ease the visualization. \textit{(right)} Boxplot of the distribution of relative changes in the number of members/followers for all accounts, for different groups (outliers are filtered).}
    \label{fig:accounts_stats}
\end{figure}

For what concerns metrics for accounts\footnote{We filtered out accounts with only 1 post to remove noise.} involved in the analysis (\emph{e.g.}, total engagement, total number of posts, number of members and total number of shared links), we report that in all cases their distribution seems to follow power laws, which are common in social networks dimensions (see Appendix, Fig. A2). When considering separately accounts who discussed on controversial topics, we observe on average an higher activity compared to the global set of accounts, but we do notice similar distributions of members; therefore, we analyze the distribution of members versus the other dimensions (see Figure~\ref{fig:accounts_stats} left, centre) and we notice that (1), as expected, accounts with a larger number of members are more active but also (2) they were more likely involved in discussions on controversial topics. These results do not change if we consider Groups or Pages alone. We also consider the relative change in the number of members/likes of accounts during the observation period. We observe that groups have larger oscillations and higher (positive) growths compared to pages (see Appendix, Fig. A3); also, we notice that accounts which discussed about controversial topics experienced a larger growth compared both to the entire set and to those which did not discuss about any of them (see Figure~\ref{fig:accounts_stats} right). However, further investigation is needed to understand whether there is a causality effect between discussing about specific topics and experiencing a growth in followers/members.

We then analyzed the total engagement generated by different accounts to understand which were the most influential in general and for each topic. In the former case in the Top-10 ranking (see Appendix, Table~\ref{tab:top10eng}) we encounter 5 pages related to newspapers, 1 to a popular pseudo-journalistic TV program (Le Iene) and 4 pages related to right-wing politicians (in particular 2 pages of Matteo Salvini and 1 page of Luca Zaia, governor of Veneto region which was one of the most affected by the virus). Each of these accounts generated from 13 to 50 M reactions during the period of observation. 

For what concerns controversial topics, we consider the Top-10 ranking  according to the total number of reactions generated only by posts related to each topic (see Appendix, Tables~\ref{tab:top105g}~\ref{tab:top10labs} and~\ref{tab:top10mig}).
For what concerns 5G, we notice that most influential accounts shared only a few related posts (from 2 to 6) but generated from 90k to almost 2 M reactions (which were accounted by 2 posts of the Italian Health Ministry).
For what concerns LABS, we notice a larger number of posts and total reactions generated, most of which are accounted by Matteo Salvini leader of the right-wing Lega party.
Finally, for what concerns MIGRANTS we see a larger number of posts/reactions w.r.t to other topics, most of which are accounted by a newspaper (``Tgcom24") and Matteo Salvini, respectively with 3.4 M and 1.1 M total engagement.

\subsection{Polarization of accounts and linguistic analysis}

\begin{figure}[!t]
    \centering
    \includegraphics[width=0.5\linewidth]{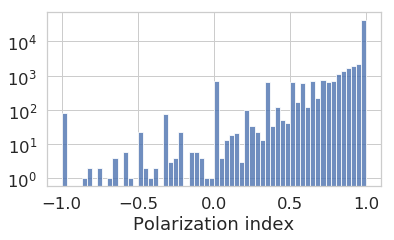}
    \caption{Histogram of the polarization scores of accounts.}
    \label{fig:pol_tot}
\end{figure}

To investigate the polarization of accounts towards different topics we introduce a polarization score $\rho = \frac{p_a-p_b}{p_a+p_b}$, where $p_b$ is the number of posts of an account about controversial topics, while $p_a$ is the number of posts of an account not about controversial topics. We define a post about controversial topics if it contains at least one of the manually selected tokens.  
The polarization index is constrained between $-1$, when all the posts of an account are about controversial topics ($p_a = 0$) and $+1$, when no posts involved controversial topics ($p_b = 0$).
Figure~\ref{fig:pol_tot} shows the distribution of the polarization scores of accounts. 
We notice a trimodal distribution: a peak at $\rho=1$ representing accounts not talking about controversial topics (the greater majority of accounts), a second peak at $\rho=0$ that includes accounts talking equally about controversial and not controversial topics, and a third lower peak at $\rho=-1$ which represents accounts posting only about controversial topics. 
In the Appendix (Figure~\ref{fig:pol_couples}), we show also how accounts are polarized when comparing topics against each other with the same rationale, i.e., by defining $p_a$ the number of posts about one controversial topic (\emph{e.g.}, \textit{5g}) and $p_b$ the number of posts about a different controversial topic (\emph{e.g.}, \textit{Labs}).
Peaks at $\rho=+1$ and $\rho=-1$ indicate that most accounts usually do not talk about more than one controversial topic. 

\begin{figure}[!t]
    \centering
    \includegraphics[width=\linewidth]{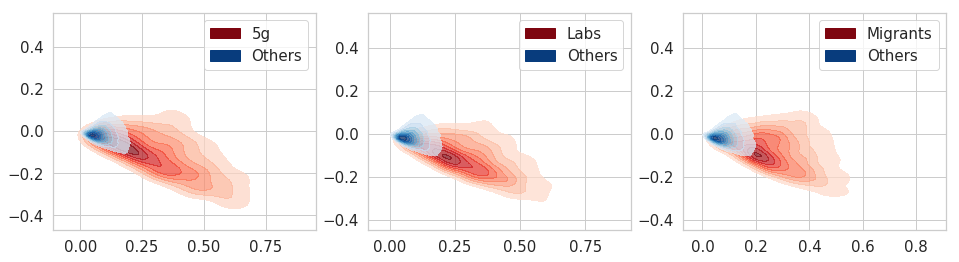}
    \caption{Distribution of the first two main components of embeddings of accounts.}
    \label{fig:KDE}
\end{figure}

In Figure~\ref{fig:KDE} we show a kernel density plot for the embedding of accounts. For visualization purposes, we select the first two PCA components. Red indicates controversial accounts whereas blue indicates remaining ones. Note that, even if the intersection of controversial accounts is negligible (see Table~\ref{tab:stats_accounts}) and the accounts are usually polarized on a single controversial topic (see Figure~\ref{fig:pol_couples}), the embeddings of the two ``classes" distribute differently in all cases. This result suggests that, since the distributions of the embeddings are different, they might be suitable input features to automatically classify controversial vs non-controversial accounts -- a task beyond the scope of this paper.

\section{Sharing diffusion network}\label{sec:network_analysis}

To better understand the patterns of sharing diffusion on Facebook, we now focus on the bipartite graph of accounts and URLs and on its projection upon the two layers, as defined in Section~\ref{sec:network_methodology}.
Since we are particularly interested in characterizing controversial URLs and accounts, we will focus on the subgraph induced by such URLs.

\subsection{Bipartite graph}\label{sec:bipartite}

By inspecting the bipartite graph we aim to investigate two related aspects of the COVID-19 infodemic on Facebook: (i) whether there are niches of accounts where (possibly extreme) conspiracy theories get diffused; and, conversely, (ii) whether there exists a (relatively) small set of accounts that, altogether, provide access to a vast majority of all available web resources.

To answer question (i), we look at the connected components of the graph.
The giant component of the entire bipartite graph includes $\approx57\%$ of all accounts, $\approx88\%$ of all urls and $\approx92\%$ of all edges.
With a bit of simplification, this means that (limited to our dataset) more than half of all Facebook accounts draw upon a unique large pool of web content -- with the remaining pieces being essentially negligible as they are fall out this giant component.
Quite interestingly, a similar scenario emerges if we only consider the set of controversial URLs: this bipartite subgraph has 55,411 vertices (18,681 accounts and 36,730 urls) and 81,707 edges, and its giant component includes $\approx72\%$ of both accounts and URLs and $\approx87\%$ of all edges.
Components other than the giant are at least two orders of magnitude smaller in both graphs.

The isolation of specific accounts and URLs into such small components seems to emerge as a consequence of (poor) marketing strategies.
For both graphs, in fact, the majority of the components consist of a small number of accounts -- often, just one -- sharing many different URLs (\emph{cf.} Figure~\ref{fig:scatter_components} in the Appendix).
Through manual investigation, we verified that such phenomenon is oftentimes caused by a website controlling one or more Facebook pages to promote its articles.
A notable case is ``howtodofor.com", which seems to use as many as 17 different accounts, none of which is apparently ascribable to the website owners.
However, a deeper and more rigorous analysis of similar cases is left to future work.

To answer question (ii), (as shown in Figure~\ref{fig:urlReach}) we observe that few accounts are sufficient to reach the majority of the URLs shared in the network. 
To reach the $25\%$ of the URLs we need 88 ($0.65\%$) accounts, to reach the $50\%$ of the URLs we need 458 ($3.42\%$) accounts, to reach the $75\%$ of the URLs we need 1,535 ($11.48\%$) accounts, finally to reach the $90\%$ of the URLs we need 3,539 ($26.48\%$) accounts.

\subsection{Controversial urls, domains and accounts}

We now consider the giant component in the projection of the controversial bipartite graph upon the two layers of URLs and accounts.
This leads to a URL graph with 26,705 vertices and 1,096,672 edges, and an account graph with 13,363 vertices and 986,509 edges.
The diameter, radius and average path length of the two graphs -- 10, 5 and 3.24 for URLs, 10, 5 and 2.86 for accounts -- depict a \emph{small world} \cite{barabasi,newman}, or even \emph{ultra-small world}\footnote{A ultra-small world has $L \propto log\,log\, N$, where $L$ is the average path length and $N$ is the number of nodes.}, further confirmed by the global efficiency -- 0.33 for URLs, 0.37 for accounts -- and the clustering coefficient -- 0.38 for URLs, 0.68 for accounts.
This means that we observe Facebook accounts which often share common sets of URLs.
At the same time, visiting a small percentage of them is enough to cover all the URLs shared in the network, as showed in Figure~\ref{fig:urlReach}.
Altogether, this suggests that if Facebook allowed users to ``jump" through groups and pages via shared URLs, they would likely get to all controversial URLs no matter the stance towards the topic.
On the one hand, finding propaganda items on Facebook appears an easy task; on the other hand, debunking articles are probably equally easy to find, if the right instruments for browsing content were provided.

\begin{figure}[!t]
    \centering
     \begin{subfigure}[b]{0.49\textwidth}
         \centering
         \includegraphics[width=\textwidth]{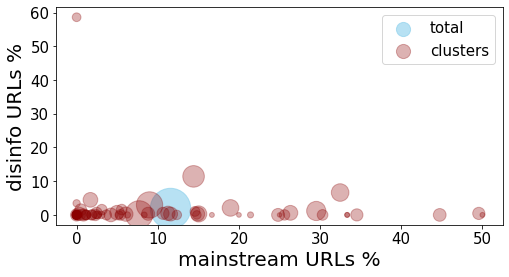}
         \caption{URL clusters}
         \label{fig:mainstream_disinfo_ratio_URLs}
    \end{subfigure}
    \hfill
    \begin{subfigure}[b]{0.49\textwidth}
         \centering
         \includegraphics[width=\textwidth]{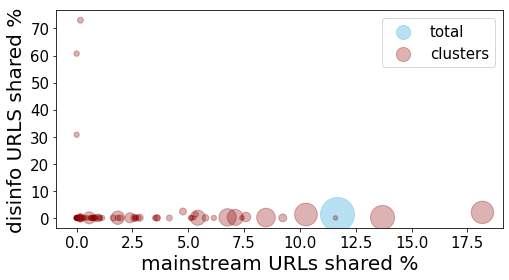}
         \caption{Account clusters}
         \label{fig:mainstream_disinfo_ratio_accounts}
     \end{subfigure}
    \caption{Prevalence of Mainstream and Disinformation domains in different clusters of the controversial URLs graph and controversial account graph.}
    \label{fig:mainstream_disinfo}
\end{figure}

To better understand whether Mainstream news and Disinformation items are well mixed on Facebook, we rely on the manual classification of sources described in Section~\ref{sec:mainstream_disinfo}.
Specifically, we aim to assess the existence of clusters of URLs and accounts with a strong prevalence of one of the two categories of domains.
As a side result, we expect to gain insights into the possibility to infer the quality of web content solely based on where and who shared the item.
We computed modularity-based clusters relying on the well-known Louvain algorithm \cite{louvain}.
In Figure~\ref{fig:mainstream_disinfo} we compare the ratio of Mainstream URLs with the ratio of Disinformation URLs present in each cluster, and in the entire graph of URLs/accounts.
We see that the ratio of deceptive URLs is very low for almost all clusters, and that in general news-related URLs are a minority.
This is mainly a consequence of the high prevalence of domains relying on user generated content and that, as such, cannot be classified -- \emph{e.g.}, ``facebook.com" and ``youtube.com" account for, respectively, $\approx48\%$ and $\approx4\%$ of all URLs.

Focusing on clusters of URLs, we argue that the most interesting cluster is cluster 2, the third largest cluster (2,179 URLs from 345 different domains) with a share of Disinformatiom domains that, albeit being barely over 10\%, is notably larger than that of the entire URL graph.
Ten domains appear in cluster 2, and they are all well-known websites which share sovereignty propaganda and false stories (from ``voxnews.info" to ``ilprimatonazionale.it").
However, the cluster also includes URLs from 20 Mainstream domains, showing that propaganda is apparently well-mixed with Mainstream information.
Conversely, the only strongly polarized cluster of URLs, which includes almost 60\% of disinfo URLs, is in reality similar to the small connected components discussed in Section~\ref{sec:bipartite} and, as such, of limited interest: it is very small (less that 100 URLs), dominated by just a few domains, and apparently consisting of a few web sites promoting each other.

For what regards accounts, almost all clusters -- including larger ones -- have a negligible presence of disinformation URLs, a further element in favor of the existence of a body of web resources shared on/by groups and pages with very different audiences.
The only clusters of accounts with a large share of disinformation are very small clusters of, at most, 12 accounts.
It is interesting to notice that the names of these clusters include a mix of references to cross-ideological political parties and local movements, thus suggesting that disinformation spreads across political affiliation.

Finally, we analyzed the centrality of individual URLs and accounts in the graph based on their PageRank\cite{pagerank}, degree and strength \cite{barabasi,newman}.
For URLs, we observe (\emph{cf.} Table~\ref{tab:urls_pagerank}) that most central URLs generally convey mainstream, institutional and scientific content, with a few notable exceptions: six Youtube videos (one of which embedded in a Facebook post) that went viral during the COVID-19 infodemic, all related to counter-information and alternative propaganda -- three of which are not available on Youtube anymore. Besides, all of these belong to the aforementioned cluster 2, which includes other widespread content as well: 2 news items from mainstream media outlets but very popular among conspirationists (an interview to an ``expert'' claiming that the virus is indeed a biological weapon, published by ``TgCom24", and an article about the role of particulate in the epidemics, published by ``AGI"); finally, a few URLs linking to scientific papers, data analysis and visualization of the pandemic.
Two insights can be drawn: on the one hand, propaganda items are so popular that they invade the general debate on Facebook and are often found in accounts that also share factual and rigorous news items; on the other hand, a categorization of disinformation domains is useful but insufficient, because most popular disinformation content is shared as user generated content on platforms such as Youtube and Facebook, thus needing manual verification to flag them as harmful.

For what concerns central accounts, we see that most of them make explicit references to anti-establishment journalists, known counter-information bloggers, and conspiracy theories.
Quite interestingly, we also see that these are mostly groups and do not coincide with the top ranking accounts emerged in Section~\ref{sec:accounts}, \emph{i.e.}, the greatest engagement is generated by accounts that are not among the most central in the network built upon URL shares.
We may argue that controversial opinions are mostly shaped on groups, based on URLs shared by other users, and then just ``gathered'' on the public pages of political leaders and parties.

\section{Conclusions}
In this paper we investigated online conversations about COVID-19 and related controversial topics on Facebook, during a period of 4 months and analyzing more than 1.5 M posts shared by almost 80k groups and pages.
We first noticed that discussions on controversial topics, which had a smaller volume of interactions compared to the pandemic in general, induced polarized clusters of accounts in terms of both topic coverage and lexicon.
We then observed that, in accordance with recent literature, sources of (supposedly) reliable information had a higher engagement compared to websites sharing unreliable content.
However, we also realized the limitations of source-based approaches when analyzing an information ecosystem wherein user generated content has a paramount role.
Finally, we highlighted a ``small-world effect'' in the sharing network of URLs, with the result that users on Facebook who navigate on a limited set of pages/groups can be potentially exposed to a wide range of content, from extreme propaganda to verified information.
In this network, the central role is taken by popular groups, in contrast with popular pages being those generating the greatest engagement.

Future directions of research include further investigating the differences in the activity of groups and pages which focus on controversial topics.
In particular, we aim to understand whether language differences might be effectively employed to distinguish accounts who were particularly active (or not) on specific subjects, and to extend the analysis of reliable \emph{vs.} unreliable information to platforms for video and image sharing such as YouTube and Instagram.

\bibliographystyle{unsrt}
\bibliography{biblio}

\newpage
\section*{Appendix}
\renewcommand{\thefigure}{A\arabic{figure}}

\setcounter{figure}{0}

The complete lists of italian words used to define the thee controversial topics are the following (we include also the feminine and plural forms, omitted here for clarity purposes). 
\begin{itemize}
\item ``5G": elettromagnetismo, onda, radiazione, wireless];
\item ``LABS": cavia, espetimento, sperimentato, sperimentazione;
\item ``MIGRANTS": africa, barcone, clandestino, extracomunitario, immigrato, islam, musulmano, negro, niger, ONG, profugo, senegal, straniero
\end{itemize}

\begin{figure}[h]
    \centering
    \includegraphics[width=\linewidth]{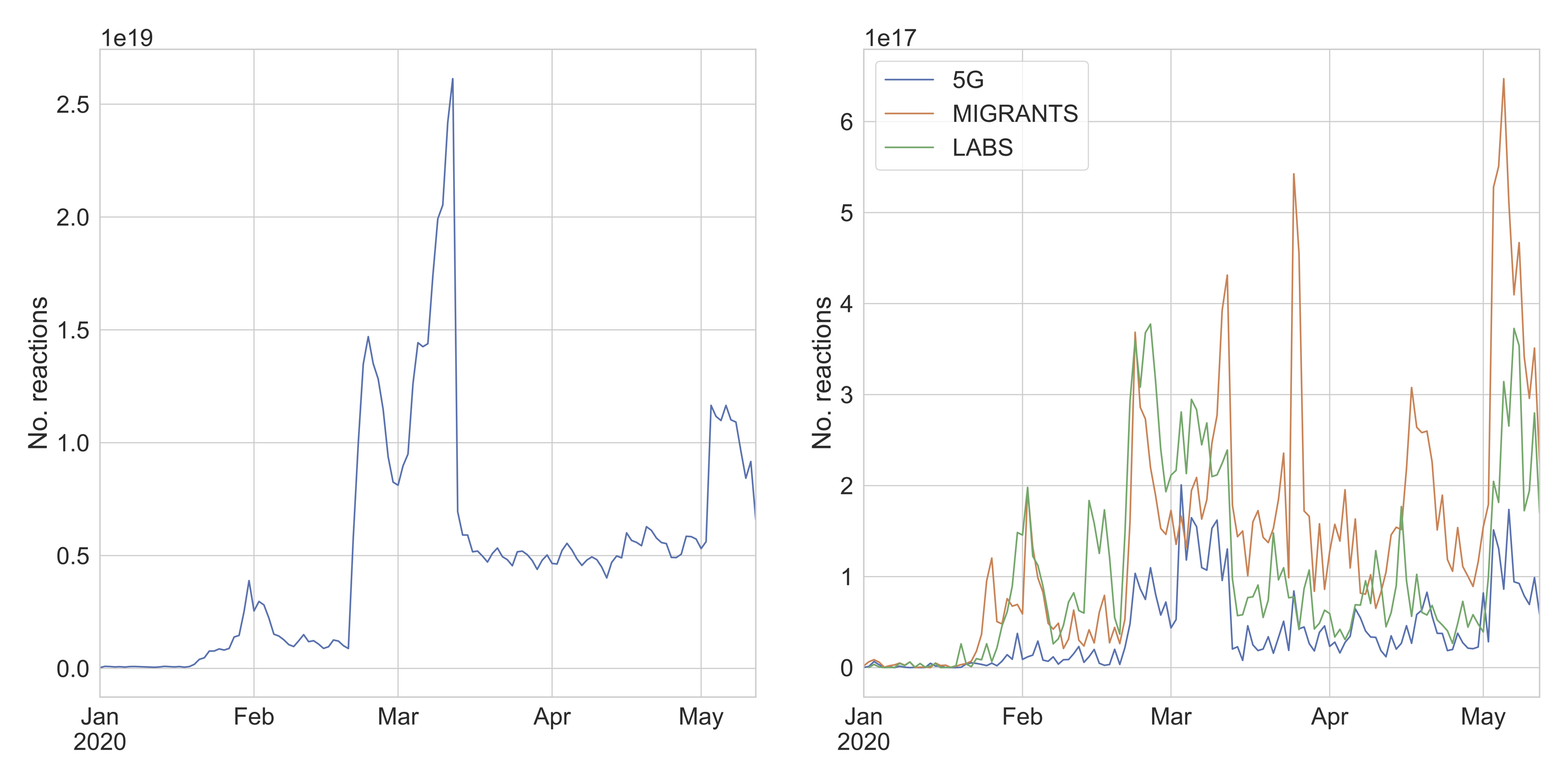}
    \caption{Time series of the daily number of reactions for all posts (left) and depending on the controversial topic (right).}
\end{figure}

\begin{table}[h]
    \centering
    \begin{tabular}{c c c}
    \textbf{Account} & \textbf{No. posts} & \textbf{No. reactions} \\
    \hline
Le Iene & 256 & 49869767\\
Fanpage.it & 3043 & 49234295\\
Corriere della Sera & 2119 & 28004758\\
Vittorio Sgarbi & 98 & 21865677\\
Tgcom24 & 2559 & 19654257\\
Sky TG24 & 1494 & 19259609\\
Notizie.it & 2513 & 16907208\\
Matteo Salvini & 215 & 16719496\\
Luca Zaia & 207 & 16068773\\
Lega - Salvini Premier & 1386 & 13187120\\
    \end{tabular}
    \caption{Top-10 ranking of all accounts by total engagement generated.}\label{tab:top10eng}
\end{table}

\begin{table}[h]
    \centering
    \begin{tabular}{c c c}
    \textbf{Account} & \textbf{No. posts} & \textbf{No. reactions} \\
    \hline
Ministero della Salute & 2 & 1863319\\
Nicola Morra & 2 & 982903\\
Che tempo che fa & 4 & 383441\\
Quarto Grado & 2 & 294430\\
Sfera & 3 & 121635\\
Lorenzo Tosa & 2 & 108586\\
Abolizione del suffragio universale & 3 & 108491\\
Il Sole 24 ORE & 6 & 104433\\
Angelo DURO & 2 & 92855\\
Fondazione Poliambulanza Istituto Ospedaliero Multispecialistico & 2 & 90481\\
    \end{tabular}
    \caption{Top-10 ranking of 5G accounts by total engagement generated.}\label{tab:top105g}
\end{table}

\begin{table}[h]
    \centering
    \begin{tabular}{c c c}
    \textbf{Account} & \textbf{No. posts} & \textbf{No. reactions} \\
    \hline
Matteo Salvini & 19 & 1443184\\
Nicola Porro & 12 & 454688\\
Silvia Sardone & 9 & 416352\\
Lega - Salvini Premier & 55 & 382654\\
Medici con l'Africa Cuamm & 20 & 331577\\
Tg3 & 18 & 280473\\
Sky TG24 & 7 & 220786\\
Local Team & 3 & 200964\\
Abolizione del suffragio universale & 7 & 173648\\
Fanpage.it & 26 & 169921\\
    \end{tabular}
    \caption{Top-10 ranking of LABS accounts by total engagement generated.}\label{tab:top10labs}
\end{table}

\begin{table}[h]
    \centering
    \begin{tabular}{c c c}
    \textbf{Account} & \textbf{No. posts} & \textbf{No. reactions} \\
    \hline
Tgcom24 & 58 & 3408934\\
Matteo Salvini & 12 & 1157440\\
Kiko.Co & 2 & 826064\\
Luca Zaia & 9 & 603693\\
Gianni Simioli & 3 & 577797\\
Vincenzo De Luca & 25 & 535635\\
Tg1 & 3 & 489346\\
Sky TG24 & 20 & 442357\\
Il Messaggero.it & 21 & 406447\\
Tg3 & 16 & 382325\\

    \end{tabular}
    \caption{Top-10 ranking of MIGRANTS accounts by total engagement generated.}\label{tab:top10mig}
\end{table}

\begin{figure}[h]
    \centering
    \includegraphics[width=0.8\linewidth]{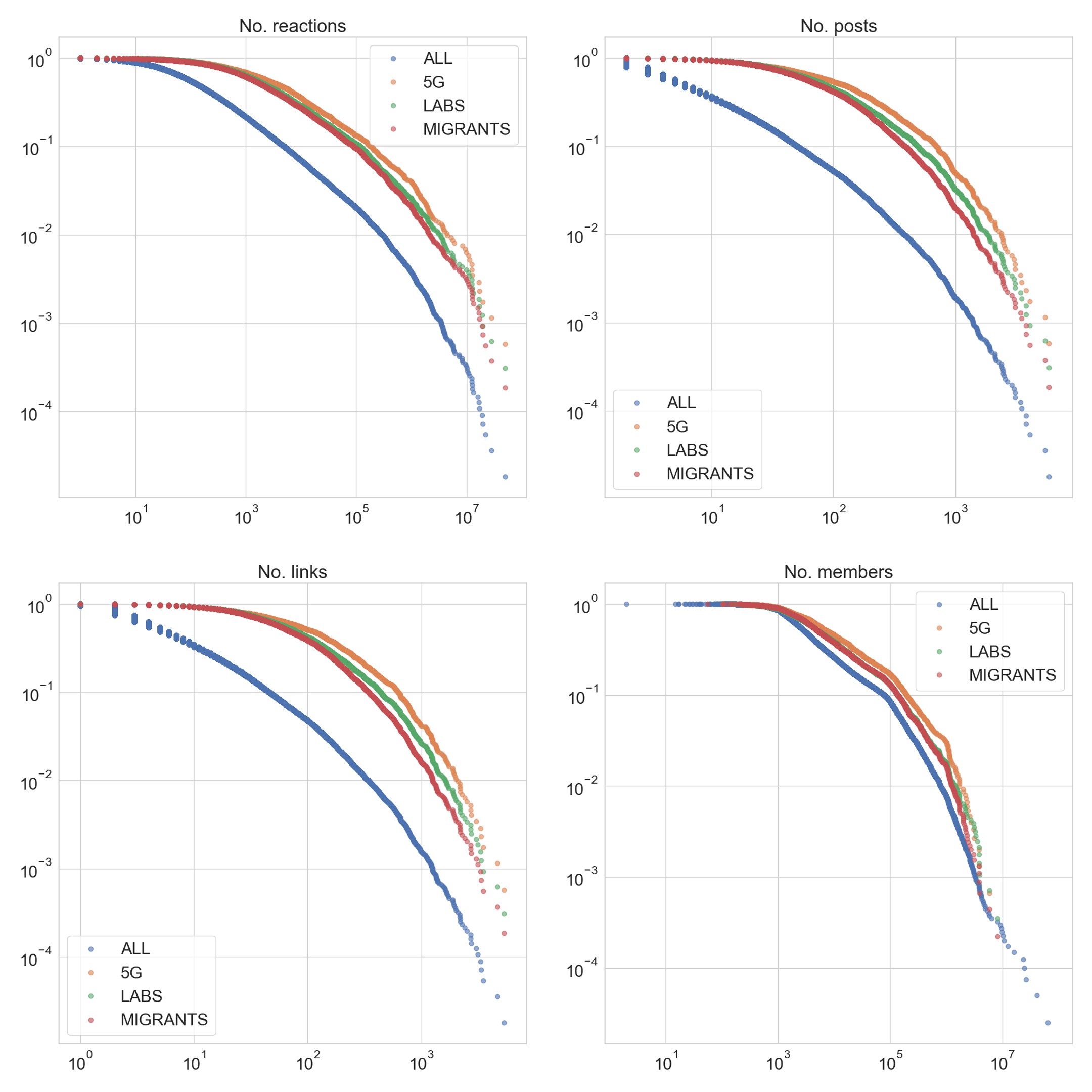}
    \caption{Complementary cumulative distribution function for several metrics. We show all accounts and according to different topics.}
\end{figure}

\begin{figure}[h]
    \centering
\includegraphics[width=0.7\linewidth]{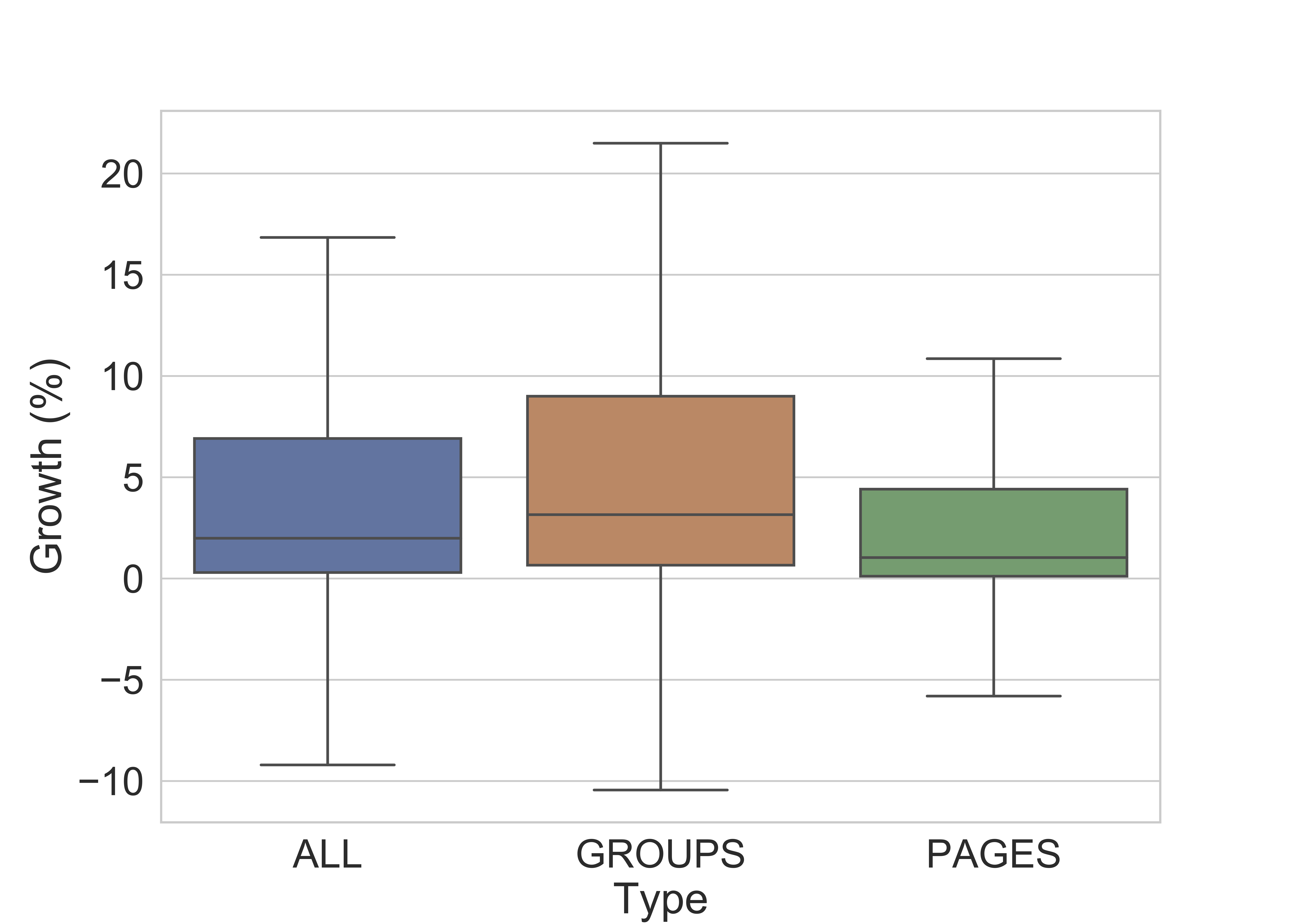}
    \caption{Boxplot of the distribution of relative changes in the number of members/followers for all accounts, groups and pages.}
\end{figure}

\begin{figure}[h]
    \centering
    \includegraphics[width=\linewidth]{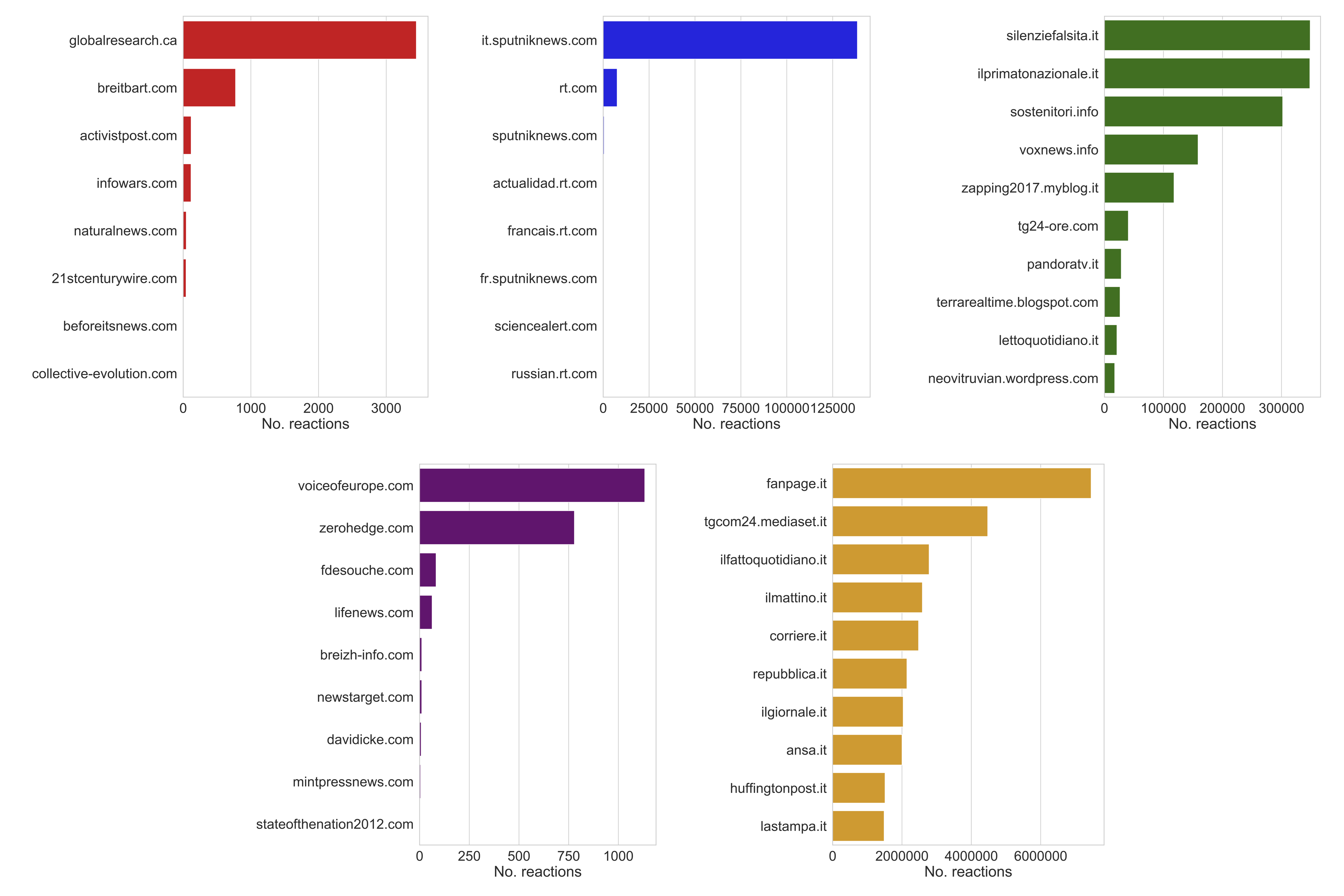}
    \caption{Top-10 ranking of news sources per different news domain according to the total engagement generated. In clockwise order from top left we show US, RU, IT, EU disinformation sources and finally IT Mainstream sources.}
\end{figure}

\begin{figure}[h]
    \centering
    \includegraphics[width=\linewidth]{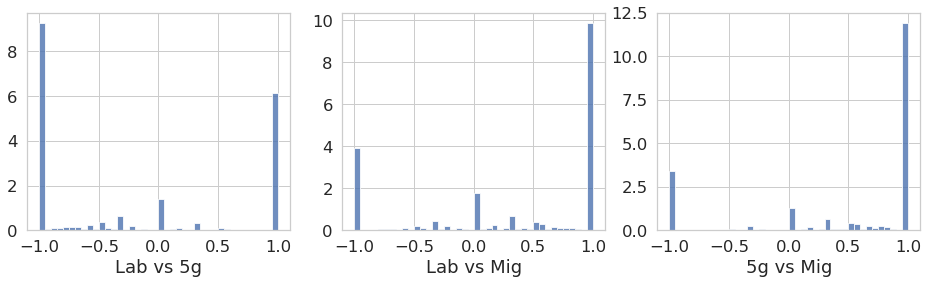}
    \caption{Normalized histogram of polarization index of accounts, by pairs of topics.}
    \label{fig:pol_couples}
\end{figure}

\begin{figure}[htbp]
    \centering
    \begin{subfigure}[b]{0.49\textwidth}
         \centering
         \includegraphics[width=\textwidth]{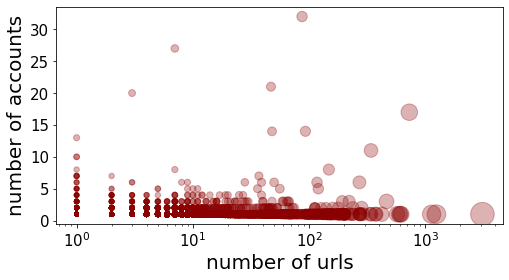}
         \caption{Whole graph}
         \label{fig:scatter_components_whole}
    \end{subfigure}
    \hfill
    \begin{subfigure}[b]{0.49\textwidth}
         \centering
         \includegraphics[width=\textwidth]{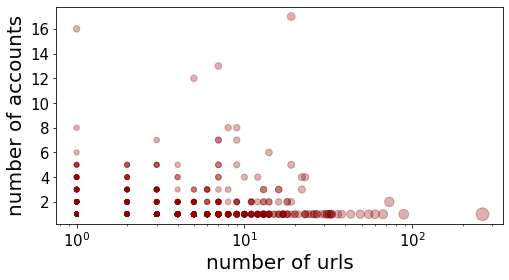}
         \caption{Controversial subgraph}
         \label{fig:scatter_components_topics}
    \end{subfigure}
    
    \caption{Number of urls and accounts in all components of the bipartite graph other than the giant, for both the whole graph and the topical subgraph. The marker size is proportional to the total number of vertices of the component.}
    \label{fig:scatter_components}
\end{figure}

\begin{figure}[!t]
    \centering
    \includegraphics[width=0.7\textwidth]{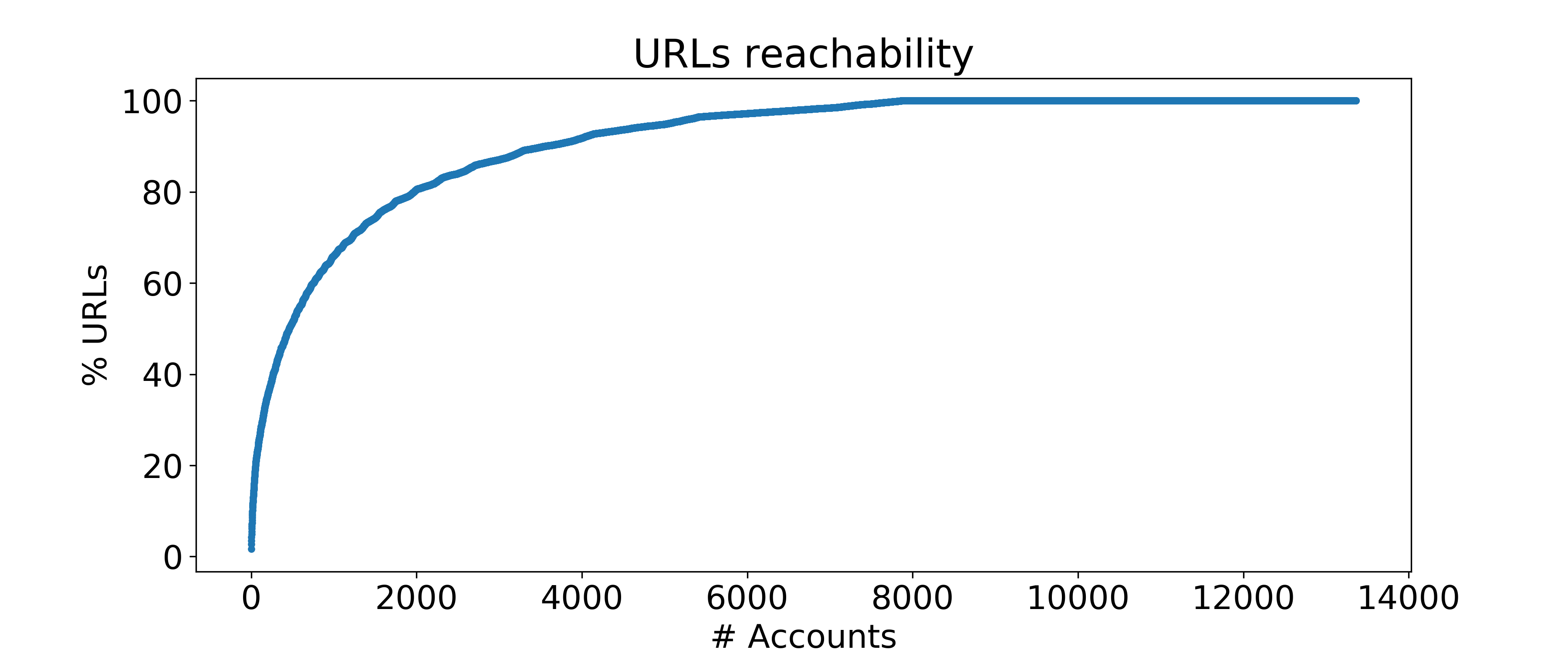}
    \caption{Percentage of reached URLs when visiting accounts in decreasing order of degree centrality.}
    \label{fig:urlReach}
\end{figure}

\begin{table}[]
    \centering
    \scriptsize
\begin{tabular}{lrrrr}
\toprule
                                               url &          pagerank &  degree &  strength &  louvain \\
\midrule
 governo.it/it/articolo/decreto-iorestoacasa-do... & 0.00211           &    2467 &      3777 &        0 \\
                   youtube.com/watch?v=RsoG7pZifTw & 0.00202           &    4605 &      8284 &        2 \\
 la7.it/piazzapulita/video/coronavirus-dentro-i... & 0.00129           &    3146 &      4351 &        0 \\
 tgcom24.mediaset.it/mondo/coronavirus-lesperto... & 0.00109           &    3320 &      4867 &        2 \\
 agi.it/cronaca/news/2020-04-24/inquinamento-pa... & 0.00103           &    2299 &      3083 &        2 \\
 salute.gov.it/portale/nuovocoronavirus/dettagl... & 0.00096           &    1586 &      2087 &        0 \\
                   youtube.com/watch?v=JXW2BNFzAtQ & 0.00087           &    2350 &      3258 &        2 \\
                     worldometers.info/coronavirus & 0.00085           &    2353 &      3052 &        2 \\
                               covidvisualizer.com & 0.00085           &    1780 &      2272 &        2 \\
                   youtube.com/watch?v=4cX7PJbV4qs & 0.0008            &    2002 &      2763 &        2 \\
            nejm.org/doi/full/10.1056/NEJMc2001468 & 0.00078           &    2504 &      3629 &        2 \\
                   youtube.com/watch?v=c8czMPZdAU8 & 0.00077           &    2145 &      3073 &        2 \\
 fanpage.it/politica/la-protezione-civile-chied... & 0.00076           &    2344 &      3242 &        2 \\
                   youtube.com/watch?v=PGVBGh32f-s & 0.00076           &    1956 &      2649 &        2 \\
 fanpage.it/attualita/coronavirus-fuga-da-milan... & 0.00074           &    2020 &      2584 &        6 \\
 it.businessinsider.com/ernesto-burgio-2-o-3-co... & 0.00074           &    1414 &      1859 &        0 \\
                    salute.gov.it/nuovocoronavirus & 0.00071           &    1020 &      1209 &        0 \\
 gisanddata.maps.arcgis.com/apps/opsdashboard/i... & 0.00068           &    1343 &      1640 &        2 \\
 facebook.com/giuseppe.provenza.5209/videos/231... & 0.00066           &    2160 &      3070 &        2 \\
\bottomrule
\end{tabular}
    \caption{The top 20 controversial URLs by pagerank.
}
    \label{tab:urls_pagerank}
\end{table}

\begin{table}[]
    \centering
    \scriptsize
\begin{tabular}{lrrrrr}
\toprule
                                           account &          pagerank &  degree &  strength &  louvain &  tot controversial \\
\midrule
                    Gruppo Tutto TRAVAGLIO Forever & 0.00169           &    3263 &      5325 &        1 &               2671 \\
                          Arcipelago delle Sardine & 0.00155           &    3041 &      4054 &        0 &               1905 \\
               Luisella   Costamagna  Fan's club & 0.00144           &    2649 &      4257 &        1 &               2119 \\
                                Dalla vostra parte & 0.00141           &    2045 &      4043 &        2 &               2397 \\
                                        Morris San & 0.00133           &    2087 &      3737 &        1 &               3492 \\
        Coronavirus Covid-19 Gruppo di discussione & 0.00121           &    2739 &      3472 &        2 &               1232 \\
                                   IO RESTO A CASA & 0.00116           &    2377 &      2862 &        5 &               5861 \\
                       Marco Travaglio\&Peter Gomez & 0.00116           &    2523 &      3495 &        1 &               1509 \\
                    Amici a cui piace Nicola Porro & 0.00113           &    1855 &      3072 &        2 &               1372 \\
 Chiediamo al Presidente Conte il test sierolog... & 0.00109           &    2015 &      2659 &        2 &                937 \\
                          Come Davide contro Golia & 0.00107           &    1781 &      3273 &        1 &               1268 \\
           Coronavirus fase 2: Italia che rinasce! & 0.00098           &    2318 &      2736 &        1 &                564 \\
      Sostenitori di Byoblu (gruppo non ufficiale) & 0.00097           &    1597 &      2672 &        1 &               1179 \\
 Vaccini Puliti. Rimozione dal commercio dei pr... & 0.00094           &    1594 &      2652 &        1 &                980 \\
                                       \#SAPEVATELO & 0.00093           &    2099 &      2676 &        0 &                941 \\
                        Con il M5S e Conte Premier & 0.00093           &    1762 &      2588 &        1 &               1204 \\
 MOVIMENTO DEI DISOCCUPATI E DEI PRECARI - LAVO... & 0.00091           &    1589 &      2543 &        1 &               1275 \\
                          Manifestazione al Senato & 0.0009            &    1727 &      2364 &        0 &               1244 \\
                           IL RUGGITO DEL CONIGLIO & 0.0009            &    2128 &      2382 &        6 &                804 \\
\bottomrule
\end{tabular}
    \caption{The top 20 controversial accounts by pagerank.
}
    \label{tab:urls_pagerank}
\end{table}

\end{document}